\documentclass[12pt]{iopart}
\usepackage{amssymb}
\begin{document}
\title{Braneworld Cosmological Perturbations in Teleparallel Gravity}

\author{A. Behboodi}%
\ead{a.behboodi@stu.umz.ac.ir} \address{Department of Physics,
Faculty of Basic Sciences,
University of Mazandaran, P. O. Box 47416-95447, Babolsar, IRAN,}%
\author{K. Nozari}%
\ead{knozari@umz.ac.ir} \address{Department of Physics, Faculty of
Basic Sciences,
University of Mazandaran, P. O. Box 47416-95447, Babolsar, IRAN,}%
\vspace{1cm}
\begin{abstract}
In this paper we find the fully gauge invariant cosmological
perturbation equations in braneworld teleparallel gravity. In this
theory, perturbations are the result of small fluctuations in the
pentad field. We derive the gauge invariant 'potentials' for both
geometric and matter variables. In teleparallel gravity, pentad
perturbations can only contain scalar and vector modes. This is in
contrast to the metric fluctuations in general relativity.
\end{abstract}

 \noindent{\it Keywords}: Teleparallel Gravity,
Cosmological Perturbations

\maketitle \vspace{2cm}
\section{Introduction and notation}
The idea that some hidden extra dimensions exist in our spacetime,
dates back to the pioneering works of Kaluza and Klein \cite{KK}.
More recently, the so called 'Braneworld' models have attracted much
attention in an attempt to explain various theoretical difficulties
such as the Hierarchy problem \cite{ADD,RS1} and the origin of
cosmological inflation \cite{Tye,DT}. One of the most interesting of
these models from a cosmological point of view, is the
Randall-Sundrum single brane model (RSII). This model has a single
positive tension brane embedded in an infinite AdS bulk. The bulk is
assumed to be empty except for a cosmological constant and ordinary
matter is confined to the brane. At low energies the gravity is
localized on the brane via the curvature of the bulk and the
standard 4D gravity is recovered on the brane, however at high
energies the gravity 'leaks' into the bulk which leads to the
modification of the standard general relativity \cite{RS2}. Various
aspects and properties of this model has been studied extensively in
the literature; for example see \cite{RSG,RSC}.

On the other hand, when studying any cosmological theory, a lot of
information can be obtained by analyzing the behavior of
cosmological perturbations and comparing it to the current
observations of the Cosmic Microwave background and large scale
structure data. This task has been done in the context of brane
cosmology by various different authors \cite{RSP}. As in the
standard 4D gravity of general relativity, studying the braneworld
cosmological perturbations is done by considering two manifolds: A
background FRW and a perturbed 'physical' manifold. Perturbation of
a quantity then defined by determining the difference between its
values in the actual physical spacetime and the background reference
manifold. The metric perturbations in GR can not be uniquely
defined, but depend on the choice of 'gauge'. A gauge essentially
can be regarded as an identification map which corresponds spacetime
points in the two manifolds. Making a different choice of gauge, may
result in the change of the values of the perturbation variables. It
may also lead to un-physical \emph{i.e.} not real perturbation
modes.

The way around this problem is to work with purely gauge invariant
equations. This means to find combinations of perturbation variables
that remain invariant under a change of the identification map. This
method, first introduced by Bardeen in \cite{Baar82},  has the
advantage of involving only real unambiguous physical quantities.
For a comprehensive review of the gauge invariant theory of
cosmological perturbations in general relativity and its
applications in inflation theory and computing the spectrum of the
cosmic microwave background radiation see references
\cite{KS84,mfb92,Durrer89,Riotto}.

General relativity, while highly successful in explaining
cosmological observations, is not the only viable theory of gravity.
More importantly some of its problems like the singularity problem
and the issue of the unification of fundamental forces led people to
try and modify or generalize it. Teleparallel theory of gravity
first introduced by Einstein \cite{Ein30} in an attempt to unify
gravity with electromagnetism. This theory in its general form is
basically the gauge theory for the translation symmetry group
\cite{Hehl07}. The theory  possesses a torsion field which can be
regarded as the translational field strength corresponding to the
coframe field \cite{Haya79}. Generally the Lagrangian of this theory
lacks Local Lorentz invariance. Restoring this symmetry mean
imposing some restrictions on the form of the Lagrangian which
results in a theory which is dynamically equivalent to general
relativity and is usually called the teleparallel equivalent of
general relativity (TEGR) in the literature \cite{Ald10,TG}.
Theleparallel gravity and its extensions \cite{LI,FT} has generated
renewed interest in recent years, in the hope that it may provide
solutions to some of the cosmological problem like the origin and
the nature of the dark energy \cite{NO1,NO2}. It should be stated
here that both teleparallel gravity and general relativity can be
regarded as special cases of a more fundamental gauge theory of
gravity called Poincare gauge theory which contains both torsion and
curvature as translational and rotational field strengths
respectively \cite{Blag}. It has been shown that a teleparallel
setup naturally arises from the low-energy effective field theory
induced on D-branes, described by a Yang-Mills theory on a flat
noncommutative space which can be regarded as the low energy limit
of string theory \cite{LSz}.

In teleparallel gravity one considers a set of $n$ linearly
independent vectors  $ e_{\i}=e^{\mu}_{~i}\partial_{\mu}$ which form
a basis in the tangent space on every point of the manifold. The
dual of this basis $\vartheta^{i}=e_{~\mu}^{i}dx^{\mu}$ are the
coframes. The dynamical variable in the TEGR theory are
$e_{~\mu}^{i}$ s, called the tetrads (or pentads in 5D) and they
relate anholonomic tangent space indices to the coordinate ones. The
spacetime metric is not an independent dynamical variable here and
is related to the tetrad through the relations

\begin{equation}
g_{\mu\nu}=\eta_{ij}e_{\mu}^{i}e_{\nu}^{j},
\end{equation}
\begin{equation}
\textbf{e}_{i}\,.\,\textbf{e}_{j}=\eta_{ij}
\end{equation}
The inverse of the tetrad is  defined by the relation
$e_{i}^{\mu}e_{\mu}^{j}=\delta_{i}^{j}.$

Teleparallel geometry, $T_{4}$ is defined by the requirement of
vanishing curvature. In the special case of TEGR the spin connection
of the theory is also assumed to be zero. This assumption is usually
called the absolute parallelism condition and by imposing it the
connection of the theory  will be the Weitzenb\"{o}ck connection
defined as \cite{Weitz23}
\begin{equation}
\Gamma^{\rho}_{\,\,\,\,\mu\nu}=e^{\rho}_{i}\partial_{\nu}e^{i}_{\mu}
\end{equation}
which unlike Livi-civita connection is not symmetric on its second
and third indices. The curvature of this connection is identically
zero and the torsion tensor is
\begin{equation}
T^{\rho}_{\,\,\,\,\mu\nu}\equiv
e_{i}^{\rho}(\partial_{\mu}e_{\nu}^{i}-\partial_{\nu}e_{\mu}^{i})\,.
\end{equation}
Contorsion tensor which denotes the difference between Livi-civita
and Weitzenb\"{o}ck connections is
\begin{equation}
K^{\mu\nu}_{\quad\rho}=-\frac{1}{2}(T^{\mu\nu}_{\quad\rho}
-T^{\nu\mu}_{\quad\rho}-T_{\rho}^{\,\,\,\,\mu\nu})
\end{equation}
and the superpotential tensor is defined as
\begin{equation}
S^{\,\,\,\,\mu\nu}_{\rho}=\frac{1}{2}(K^{\mu\nu}_{\quad\rho}
+\delta^{\mu}_{\rho}T^{\alpha\nu}_{\quad\alpha}-\delta^{\nu}_{\rho}
T^{\alpha\mu}_{\quad\alpha})\,.
\end{equation}
In correspondence with Ricci scalar, one can define torsion scalar
\begin{equation}
T=S^{\,\,\,\,\mu\nu}_{\rho}T^{\rho}_{\,\,\,\,\mu\nu}
\end{equation}

Recently, braneworld models in the context of teleparallel gravity
has been studied and its setup and fundamental equations has been
derived \cite{BAN1,BAN2}. It is understood that the RS setup in TEGR
is dynamically different from General Relativity. This fact has
motivated the authors to study the dynamics of cosmological
perturbations in this setup. In this paper, by perturbing the
coframe components (or pentads in 5D), we present the fully gauge
invariant cosmological perturbations equations in braneworld
teleparallel gravity. Unlike the metric perturbation in general
relativity which contains scalar, vector and tensor modes, pentads
can only contain up to vector perturbations. The tensor part of the
perturbation spectrum in teleparallel gravity originates directly
from the contorsion tensor, which is the difference between
Levi-Civita connection of general relativity and Weitzenb\"{o}ck
connection of teleparallel gravity. As stated, to make the
perturbation equations gauge invariant, one must find gauge
invariant combinations of perturbation variables and rewrite the
equations in terms of these parameters. These combinations in
braneworld teleparallel gravity, in analogy to the Bardeen
potentials of general relativity, has been derived for both
geometric and matter perturbations in the following sections. 5D
models based on teleparallel gravity are also studied in Refs.
\cite{BNO,GLT}.

Except when specifically stated, the notation we use throughout this
paper is as follows:  Lower case Latin letters  $a,b,...,i,j,...$
run over $1,2,3$ and label spatial tangent space coordinates. The
Greek indices $\alpha, \beta,...,\mu,\nu,..$ run over $1,2,3$ and
refer to the spatial spacetime coordinates. The letter $I$ refers to
5D tangent space indices. The rest of the upper case Latin Letters
$M,N,R,L$ refer to 5D coordinate space indices.

\section{Geometry Perturbations}
So far no trace of fifth dimension has been observed on the brane,
therefore one can consider it separate from other dimensions. This
will give the 5D spacetime, the layered structure \cite{Blag}.
Therefore the 5D tangent coordinate system
$\hat{e}_{I}=(\hat{e}_{i},\hat{e}_{\bar{5}})$ can be written as

\begin{equation}
\hat{e}_{M}=e^{i}_{\,\,M}\hat{e}_{i}+e^{\bar{5}}_{\,\,M}\hat{e}_{\bar{5}}
\end{equation}
Here the first four vectors define the coordinate system on the
layer and the fifth dimension can be chosen to be normal to the
other four dimensions. The above orthogonality condition leads us to
the point that the expansion of $\hat{e}_{\bar{5}}$ cannot contain
any of 4D $\hat{e}_{i}$ vectors. Then we have

\begin{equation}
\hat{e}_{\mu}=e^{i}_{\,\,\mu}\hat{e}_{i}+e^{\bar{5}}_{\,\,\mu}\hat{e}_{\bar{5}}\qquad
\hat{e}_{\bar{5}}=e^{\bar{5}}_{\,\,5}\hat{e}_{\bar{5}}
\end{equation}
So the general form of pentad and inverse pentad respectively are
\[e_{\,\,M}^{I}=\Bigg(\matrix{ e_{\,\,\mu}^{i} & 0\cr e_{\,\,\mu}^{\bar{5}} &
e_{5}^{\bar{5}}}\Bigg)\]
 and
\[h_{I}^{\,\,M}=\Bigg(\matrix{ h^{\,\,\mu}_{i} & h^{\,\,\mu}_{\bar{5}}\cr 0 &
h_{\bar{5}}^{5}}\Bigg)\]

Where $e_{5}^{\bar{5}}e_{5}^{\bar{5}}=1$ and
$e_{\,\,\mu}^{i}h^{\,\,5}_{i}+e_{\,\,\mu}^{\bar{5}}h_{5}^{\bar{5}}=0$
are also satisfied. The capital Latin letters indices are 5
dimensional and small Latin and Greek letter run as 0,1,2,3.\\
Considering the 5D Anti de sitter background and a 4D FRW brane with
flat spatial part, The nonzero unperturbed pentads will be
\begin{equation}
e^{\overline{0}}_{0}=n,\qquad e^{\overline{5}}_{5}=1,\qquad
e^{i}_{\alpha}=a\delta^{i}_{\alpha}
\end{equation}
Here the Latin and Greek indices are spatial and the overline
indices refer to the tangent space coordinates. The nonzero
unperturbed Weitzenb\"{o}ck connection are
\begin{equation}
\Gamma^{0}_{\,\,00}=\frac{\dot{n}}{n},\qquad
\Gamma^{\alpha}_{\,\,\beta0}=\frac{\dot{a}}{a}\delta^{\alpha}_{\beta},\qquad
\Gamma^{0}_{\,\,05}=\frac{n'}{n},\qquad
\Gamma^{\alpha}_{\,\,\beta5}=\frac{a'}{a}\delta^{\alpha}_{\beta}
\end{equation}
 To perturb the geometry of the case, one needs to perturb the pentads. Unlike
general relativity which has metrics as dynamical variable, here
pentads are our dynamical variables. By using the orthogonality
condition $(8)$, the general form of the perturbed FRW pentad can be
written as

\begin{eqnarray}
e^{\bar{0}}_{\,\,0}&=&n(1+A)~~~~,~~~~~
e^{\bar{0}}_{\,\,\mu}=aC_{\mu}~~~~~,~~~~~
e^{i}_{\,\,0}=aB^{i}~~~~~,~~~~~ e^{\bar{0}}_{\,\,5}=0\nonumber\\
e^{\bar{5}}_{\,\,0}&=&n\alpha~~~,~~~
e^{\bar{5}}_{\,\,\mu}=aF_{\mu}~~~,~~~ e^{i}_{\,\,5}=0~~~,~~~
e^{\bar{5}}_{\,\,5}=1+\varphi~~~,~~~ e^{i}_{\,\,\mu}=a(\delta^{i}_{\,\,\mu}+D^{i}_{\,\,\mu})\nonumber\\
\end{eqnarray}
And the inverse pentads are
\begin{eqnarray}
h^{\,\,0}_{\bar{0}}=\frac{1}{n}(1-A)~~~,~~~
h^{\,\,\mu}_{\bar{0}}=-\frac{a}{n}B^{\mu}~~~,~~~
h^{\,\,0}_{i}=-\frac{a}{n}C_{i}~~~,~~~ h^{\,\,5}_{\bar{0}}=-\alpha~~~,~~~ h^{\,\,0}_{\bar{5}}&=&0\nonumber\\
h^{\,\,\mu}_{\bar{5}}=0~~~,~~~ h^{\,\,5}_{i}=-F_{i}~~~,~~~
h^{\,\,5}_{\bar{5}}=1-\varphi~~~,~~~ h^{\,\,\mu}_{i}=\frac{1}{a}(\delta^{\mu}_{i}-D^{\mu}_{\,\,\,i})\nonumber\\
\end{eqnarray}
 Where $A$ , $\varphi$ and $\alpha$ are scalar
perturbations and $C_{\mu}$, $B^{i}$, $F_{\mu}$, and $D^{i}_{\,\mu}$
are vector perturbations and along with scalar ones depend on
coordinates ($\tau$, $x^{\alpha}$, $y$). $\tau$ is the conformal
time parameter and $y$ is the coordinate of fifth dimension. Note
that here we don't have tensorial perturbations since pentad unlike
metric is a vector and cannot be perturbed tonsorially. The torsion,
Contorsion, superpotential and torsion scalar of the above pentad is
given in appendix A. The calculations has been done up to the first
order of
perturbations.\\

 To have the gauge invariant geometry,
 the Lie derivative of the added perturbation fields should stay invariant under following
coordinate transformations

\begin{equation}
x^{\mu}\rightarrow \tilde{x}^{\mu}=x^{\mu}+\xi^{\mu}
~~~,~~~\xi^{a}=(T,L^{\alpha},Y)
\end{equation}
In the language of teleparallel gravity, the Lie derivative of the
pentad is
\begin{equation}
\Delta \delta e_{~M}^{I}=e_{~N}^{I}\nabla_{M} \xi^{N}
\end{equation}
This important result was first derived in \cite{Stewart90} (see
also \cite{SW}) and is usually called the 'Stewart Lemma' in the
literature.

Using the mentioned Stewart lemma, we can calculate the change in
the pentad perturbation parameters under the gauge transformation as

\begin{eqnarray}
\Delta A&=&\dot{T}+\frac{\dot{n}}{n}T\\
\Delta C_{\mu}&=&\frac{n}{a}D_{\mu}T\\
\Delta B^{i}&=&\dot{L}^{i}+\frac{\dot{a}}{a}L^{i}\\
\Delta F_{\mu}&=&\frac{1}{a}D_{\mu}Y\\
\Delta D_{~\mu}^{i}&=&D_{\mu}L^{i}\\
\Delta \Phi&=&Y'\\
\Delta \alpha&=&\frac{1}{n}\dot{Y}
\end{eqnarray}
Where a dot denotes differentiation with respect to proper time and
a prime denotes differentiation with respect to fifth dimension. To
consider only the scalar perturbations, we define the scalar part of
the vector perturbations as
\begin{equation}
B^{i}=D^{i}B  \quad   C_{\mu}=D_{\mu}C  \quad F_{\mu}=D_{\mu}F \quad
D_{\mu}^{\mu}=D
\end{equation}
Then the transformations of scalar perturbations are
\begin{eqnarray}
A&\rightarrow&A+\dot{T}+\frac{\dot{n}}{n}T\\
B&\rightarrow&B+\dot{L}+\frac{\dot{a}}{a}L\\
C&\rightarrow&C+\frac{n}{a}T\\
F&\rightarrow&F+\frac{1}{a}Y\\
D&\rightarrow&D+L\\
\Phi &\rightarrow& \Phi+Y'\\
\alpha&\rightarrow&\alpha+\frac{1}{n}\dot{Y}
\end{eqnarray}
It turns out then that the theory is not gauge invariant. To make it
invariant we repeat Bardeen's method. We define new fields out of
above old ones in a way that they stay invariant under $(14)$.
\\
\begin{eqnarray}
\Psi_{1}&=& B-\dot{D}-\frac{\dot{a}}{a}D\\
\Psi_{2}&=& A-\frac{a}{n}\dot{C}-\frac{\dot{a}}{n}C\\
\Psi_{3}&=&\Phi-a'F-a F'\\
\Psi_{4}&=&\alpha-\frac{a}{n}\dot{F}+\frac{\dot{a}}{n}F
\end{eqnarray}

These geometric objects, built from the pentad perturbation
variables, clearly remain invariant under the gauge transformation
$(16)$. Comparing with the obtained potential in 5D RS setup in
general relativity, one can easily find the above potentials are
different from the ones derived in general relativity \cite{RSP}.

\section{Matter Perturbation}
We repeat the same calculations for the matter perturbations. In RS
model of brane-world, matter is confined to the brane and bulk can
be chosen to contain only a 5D cosmological constant. We consider
the matter on the brane to be a perfect fluid ($-\rho,p,p,p$).\\
We can generally perturb the energy-momentum as
\begin{equation}
\delta T^{0}_{0}=-\rho\delta\rho ,\qquad \delta T^{0}_{\alpha}=
q_{\alpha} ,\qquad T^{\beta}_{\alpha}=p\delta p\delta
^{\beta}_{\alpha}+p\pi^{\beta}_{\alpha}
\end{equation}
Where $\delta\rho$ and $\delta p$ are scalar, $q_{\alpha}$ vector
and  $\pi^{\beta}_{\alpha}$ is purely tensor (
$\pi^{\alpha}_{\,\,\alpha}=0$).\\
Under a 4D coordinate transformation, the perturbed energy-momentum
tensor will transform like
\begin{eqnarray}
\delta \rho&\rightarrow& \delta
\rho+\hat{\rho}T+2\dot{T}\\
q_{\alpha}&\rightarrow& q_{\alpha}+p(\dot{L}_{\alpha}+\frac{\dot{a}}{a}L_{\alpha})-\rho D_{\alpha}T\\
\delta p&\rightarrow& \delta
p+\hat{p}T-\frac{\dot{a}}{a}T+\frac{2}{3}D_{\alpha}L^{\alpha}\\
\pi_{\alpha\beta}&\rightarrow&
\pi_{\alpha\beta}+2D_{(\alpha}L_{\beta)}-\frac{2}{3}\gamma_{\alpha\beta}D_{\delta}L^{\delta}
\end{eqnarray}
Considering again only the scalar parts, we have $q_{\alpha}=
D_{\alpha}q$ , $\pi_{\alpha\beta}=\Delta_{\alpha\beta}\pi$ and
$L_{\alpha}=D_{\alpha}L$ then
\\
\begin{eqnarray}
\delta \rho&\rightarrow& \delta
\rho+\hat{\rho}T+2\dot{T}\\\
q&\rightarrow& q+p(\dot{L}+\frac{\dot{a}}{a}L)-\rho T\\
\delta p&\rightarrow& \delta
p+\hat{p}T-\frac{\dot{a}}{a}T+\frac{2}{3}\Delta L\\
\pi &\rightarrow& \pi+2L
\end{eqnarray}
Where $(\hat{\rho}$ denotes $\frac{\dot{\rho}}{\rho}$ and $\hat{p}$
denotes $\frac{\dot{p}}{p}$. Similar to geometry potentials one can
make matter potential in a way that they stay invariant under above
transformations. The gauge invariant matter potentials then are
\begin{eqnarray}
\Psi_{\rho}&=&\delta\rho-2A-\frac{a}{n}(\hat{\rho}-2\frac{\dot{n}}{n})C\\
\Psi_{q}&=&q-p B+\frac{a}{n}\rho C\\
\Psi_{p}&=&\delta
p-\frac{a}{n}(\hat{p}-\frac{\dot{a}}{a})C-\frac{2}{3}\Delta D\\
\Psi_{\pi}&=&\pi-2D
\end{eqnarray}
\\
\section{Gauge invariant 5D field equations}
The gravitational action in TEGR is
\begin{equation}
I=\frac{1}{16\pi G}\int d^{5}x\,|e|\, T
\end{equation}
Where $|e|$ is the determinant of the pentad $e^{I}_{M}$. The field
equations then can be obtained by variation of (48) with respect to
the pentad
\begin{equation}
e^{-1}\partial_{M}(ee_{I}^{\,\,R}S^{\,\,\,\,MN}_{R})
-e_{I}^{\,\,N}T_{\,\,\,\,ML}^{R}S^{\,\,\,\,LM}_{R}
+\frac{1}{4}e_{I}^{\,\,N}T=4\pi G
e_{I}^{R}(-\Lambda_{5}+\delta(y)S_{R}^{\,\,N})
\end{equation}
Where $\Lambda_{5}$ is the bulk cosmological constant. Here $I$
refers to 5D tangent space and the rest refer to 5D coordinate space
and $S_{R}^{\,\,N}=(-\rho,p,p,p,0)$ is the matter which is localized
on the brane. Various components of the left-hand side of the
perturbed 5D teleparallel field equation $(49)$ is given in appendix
B.

Brane is a 4D hypersurface which divides the 5D bulk in two regions.
The two sides of this hypersurface is connected via the junction
conditions. The junction conditions for the TEGR set-up of RS model
has been derived in \cite{BAN1} as
\begin{equation}
e_{I}^{\,\,\,R}\,[S^{\,\,\,\,MN}_{R}]\,n_{M}=4\pi
G\,\,S_{I}^{\,\,\,N}
\end{equation}
where $n_{M}$ is the unit vector normal to the brane and can be
chosen
to be $(0,0,0,0,1)$. \\
writing (50) in components, we have
\begin{eqnarray}
(I=0, N=0) :
-\frac{3a'}{an}(1-2\Phi-A)-\frac{1}{n}D_{~~\beta}^{'\beta}=4\pi
G\frac{\rho}{n^{3}}(1+\delta\rho-3A)
\end{eqnarray}
\\
\begin{eqnarray}
(I=0, N=\alpha)
&:&(\frac{a'}{2a}+\frac{n'}{2an}-\frac{a'}{a^{2}})C^{\alpha}+(\frac{2a'}{n}+\frac{an'}{2n^{2}}+\frac{a'}{an}-\frac{n'}{2n^{2}})B^{\alpha}
\nonumber\\
&~&+\frac{1}{2n}B^{'\alpha}-\frac{1}{2a}C^{'\alpha}+\frac{\dot{a}}{2na^{2}}F^{\alpha}
+\frac{1}{2na}\dot{F}^{\alpha}-\frac{1}{2a^{2}}\partial^{\alpha}\alpha\nonumber\\
&~& =4\pi
G\Big(-(\frac{P}{an}+\frac{\rho}{n^{3}})B^{\alpha}+\frac{\rho}{an^{2}}C^{\alpha}+\frac{1}{na^{2}}q^{\alpha}\Big)
\end{eqnarray}
\\
\begin{eqnarray}
(I=a,
N=0)&:&-\frac{1}{2an}\partial_{a}\alpha+\frac{5\dot{a}}{2an^{2}}F_{a}+\frac{1}{2n^{2}}\dot{F}_{a}+(\frac{n'}{2n^{2}}+\frac{5a'}{2n})C_{a}
+\frac{1}{2n}C'_{a}\nonumber\\
&~&\frac{n'a}{2n^{3}}(a-1)B_{a}-\frac{a}{2n^{2}}B'_{a}=4\pi
G\Big(\frac{P}{an^{2}}B_{a}-(\frac{p}{a^{2}n}+\frac{\rho
a}{n^{3}})C_{a}
-\frac{1}{an^{2}}q_{a}\Big)\nonumber\\
\end{eqnarray}
\\
\begin{eqnarray}
(I=a, N=\alpha)
&:&-\frac{1}{a}\partial^{[\alpha}F_{a]}+\frac{1}{a}(D_{~~~a)}^{'(\alpha}-\delta^{\alpha}_{a}D_{~~\beta}^{'\beta})
+\frac{1}{a}(\frac{2a'}{a}+\frac{n'}{n})D^{\alpha}_{~a}\nonumber\\
&~&-\frac{1}{a}A'\delta^{\alpha}_{a}+(\frac{4a'}{a^{2}}+\frac{2n'}{an})\Phi\delta^{\alpha}_{a}
-\frac{2\dot{a}}{a^{2}n}\alpha\delta^{\alpha}_{a}-(\frac{2a'}{a^{2}}+\frac{n'}{an})\delta^{\alpha}_{a}\nonumber\\
&~&=4\pi G\frac{1}{a^{3}}\Big(p(1+\delta
p)\delta^{\alpha}_{a}+p\pi^{~\alpha}_{a}-p(D^{~\alpha}_{a}+2D^{\alpha}_{~a})\Big)
\end{eqnarray}
\\
\begin{eqnarray}
(I=5, N=0)
:\frac{1}{n^{2}}\Big(\partial_{\alpha}B^{\alpha}-\partial_{0}D_{~\alpha}^{\alpha}\Big)-\frac{3\dot{a}}{an^{2}}(1-2A-\Phi)
+\frac{3a'}{an}\alpha=0\nonumber\\
\end{eqnarray}
\\
\begin{eqnarray}
(I=5,
N=\alpha)&:&-\frac{1}{a^{2}}\partial^{\alpha}A-\frac{2}{a^{2}}\partial^{[\alpha}D_{~~\beta]}^{\beta}-\frac{3\dot{a}}{an^{2}}
B^{\alpha}+(\frac{4\dot{a}}{a^{2}n}-\frac{\dot{a}}{an})C^{\alpha}\nonumber\\
&~&+\frac{1}{na}\dot{C}^{\alpha}+(\frac{4a'}{a^{2}}+\frac{n'}{an})F^{\alpha}-\frac{2}{a^{2}}\partial^{\alpha}\Phi+\frac{2}{a}F^{'\alpha}=0\nonumber\\
\end{eqnarray}
Note that the left hand sides of the above equations should be
evaluated at the position of the brane. Using the background field
equations and the definition of geometric and matter potentials
$(31-34)$ and $(44-47)$, we can rewrite various components of the
field equation $(49)$ in terms of gauge invariant variables. The
result for the fully Gauge invariant field
 equations are as follows\\
\begin{eqnarray}
(I=a, N=\alpha)
&:&\frac{1}{a}\Big[\partial_{a}\partial^{\alpha}(\frac{1}{a^{2}}\Psi_{1}+\frac{2\dot{a}}{an^{2}}\Psi_{2})-\nabla^{2}(\frac{1}{a^{2}}\Psi_{1}
+\frac{2\dot{a}}{an^{2}}\Psi_{2})\delta^{\alpha}_{a})\Big]\nonumber\\
&~&+\frac{1}{2n^{3}a}\dot{\Psi}_{1}\delta^{\alpha}_{a}-
\frac{\dot{a}n'}{a^{2}n^{2}}\Psi_{4}\delta^{\alpha}_{a}-\frac{4a'n'}{a^{2}n}\Psi_{3}\delta^{\alpha}_{a}
-\frac{2\dot{a}}{a^{2}n^{3}}\dot{\Psi}_{2}\delta^{\alpha}_{a}\nonumber\\
&~&+\frac{\dot{n}}{2n^{4}}\Big((\ddot{\Psi}_{1}
+\ddot{\Psi}_{3})\delta^{\alpha}_{a}\Big)+\frac{1}{2a}\Big((\Psi''_{1}+\Psi''_{2})\delta^{\alpha}_{a}\Big)\nonumber\\
&~&\frac{a'}{a^{2}}\Big((\Psi'_{1}+\Psi'_{2})\delta^{\alpha}_{a}\Big)+\frac{a'}{a^{2}n}\Psi'_{3}\delta^{\alpha}_{a}-\frac{\dot{a}}{a^{2}n}\Psi'_{4}\delta^{\alpha}_{a}=0
\end{eqnarray}
\\
\begin{eqnarray}
(I=a,
N=5)&:&\frac{\dot{a}}{2a^{2}n}\partial_{a}\dot{\Psi}_{4}-\frac{3\dot{a}}{2a^{2}}\partial_{a}\Psi_{2}
-\frac{3}{2}\frac{a'}{a^{2}}\partial_{a}\Psi_{1}+(\frac{3a'n'}{4n^{3}}+\frac{a\dot{n}'}{4n^{3}}
-\frac{a\dot{n}n'}{2n^{2}})\partial_{a}\dot{\Psi_{1}}\nonumber\\
&~&+(-\frac{5a^{'2}}{2a^{3}}-\frac{2n'a'}{a^{2}n}+\frac{3\dot{a}^{2}}{2a^{2}n^{2}})\partial_{a}\Psi_{4}=0
\end{eqnarray}
\\
\begin{eqnarray}
(I=5, N=\alpha)&:&
-(\frac{a'}{a^{2}}+\frac{n'}{a^{2}n})\partial^{\alpha}\Psi_{3}+(\frac{a'}{2a^{2}}+\frac{n'}{2a^{2}n})\partial^{\alpha}\Psi_{2}
\nonumber\\
&~&-\frac{1}{a^{2}}\partial^{\alpha}\Psi'_{3}+\frac{1}{2a^{2}}\partial^{\alpha}\Psi'_{2}
+\frac{1}{4a^{2}n}\partial^{\alpha}\dot{\Psi}_{4}+\frac{\dot{a}}{a^{3}n}\partial^{\alpha}\Psi_{4}=0
\end{eqnarray}
\\
\begin{eqnarray}
(I=5, N=5)
&:&\frac{1}{a^{2}}\nabla^{2}(\Psi_{3}+\frac{1}{2}\Psi_{2})+(\frac{\dot{n}}{an^{3}}-\frac{3\dot{a}}{2an^{2}})\dot{\Psi}_{1}
+\frac{3\dot{a}^{2}}{2a^{2}n^{2}}\Psi_{2}=0\nonumber\\
\end{eqnarray}
\\
\begin{eqnarray}
(I=a,
N=0)&:&\frac{a^{2}}{2n}\partial_{a}\Psi_{1}+(\frac{a'a}{2}+\frac{n'a^{2}}{2n})\partial_{a}\Psi_{4}-\frac{a'}{2a^{2}n}\Psi'_{4}
\nonumber\\
&~&+\frac{3\dot{a}}{2a^{2}n}\partial_{a}\Psi_{3}+\frac{a\dot{a}}{n}\partial_{a}\Psi_{1}=0
\end{eqnarray}
\\
\begin{eqnarray}
(I=5, N=0)
:-\frac{1}{4a^{2}n}\nabla^{2}\Psi_{4}-\frac{\dot{a}}{an^{2}}\Psi_{3}+\frac{1}{2an^{2}}\nabla^{2}\Psi_{q}
+(\frac{3a'}{2n^{2}}+\frac{2n'}{2n^{3}})\Psi'_{1}=0\nonumber\\
\end{eqnarray}
\\
\begin{eqnarray}
(I=0,
N=0)&:&(\frac{5\dot{a}}{4a^{2}n^{2}}-\frac{\dot{a}}{2an^{3}})\nabla^{2}\Psi_{2}
-\frac{2\dot{a}}{an^{3}}\Psi_{2}+3(\frac{a^{'2}}{a^{2}n}+\frac{2a'n'}{an^{2}})\Psi_{3}\nonumber\\
&~&+(\frac{\dot{a}n'}{an^{3}}-\frac{3a'\dot{a}}{a^{2}n^{2}}-\frac{3n'\dot{a}}{2an^{2}})\Psi_{4}+\frac{3a'n'}{2na^{2}}\Psi_{1}
+(\frac{1+3a'}{na})\Psi'_{1}=0\nonumber\\
\end{eqnarray}
\\
\begin{eqnarray}
(I=0, N=\alpha)
&:&\frac{1}{na^{3}}\partial^{\alpha}(\Psi'_{3}+(\frac{n'}{a^{2}n}+\frac{a'}{a^{3}})\partial^{\alpha}\Psi_{3}-\frac{n'}{2an^{2}}\partial^{\alpha}\Psi_{4}
-\frac{1}{2a^{2}}\partial^{\alpha}\Psi'_{2}\nonumber\\
&~&-(\frac{n'}{2a^{2}n}+\frac{a'}{2a^{3}})\partial^{\alpha}\Psi_{2}
=0
\end{eqnarray}
The first four of these equations are dynamical equations and the
rest act as constraints. This set of equations can be solved for
four potentials $\Psi_{1}$, $\Psi_{2}$, $\Psi_{3}$ and $\Psi_{4}$ in
the bulk subject to boundary conditions provided by the junction
conditions (eqs (51-56)). The resulting perturbation dynamics is in
general different from general relativity. One can also bring the
junction conditions into gauge invariant form. The results are as
follows\\
\begin{eqnarray}
(I=0, N=0)
:\frac{3a'}{an}\Psi'_{1}+\frac{6a'}{an}\Psi_{3}+\frac{a'\dot{n}}{an^{2}}\Psi_{2}=4\pi
G\frac{1}{n^{3}}\Psi_{\rho}
\end{eqnarray}
\\
\begin{eqnarray}
(I=0,
N=\alpha):-\frac{1}{2a^{2}}\partial^{\alpha}\Psi_{4}+\frac{n'}{2na
}\partial_{a}\Psi_{1}=4\pi G \frac{1}{na^{2}}\Psi_{q}
\end{eqnarray}
\\
\begin{eqnarray}
(I=a,
N=0)&:&\frac{1}{2an}\partial_{a}\Psi_{4}+\frac{3a^{2}n'}{2n^{3}}\partial_{a}\Psi_{2}
=\frac{1}{an^{2}}\partial_{a}\Psi_{q}
\end{eqnarray}
\\
\begin{eqnarray}
(I=a,
N=\alpha)&:&\frac{1}{a}\partial_{a}\Psi'_{1}\delta^{\alpha}_{a}-\frac{2\dot{a}}{a^{2}n}\Psi_{4}\delta^{\alpha}_{a}
\frac{1}{a}\partial_{a}\Psi'_{2}+\frac{4a'}{a^{2}}\partial_{a}\Psi_{3}\nonumber\\
&~&=4\pi G \frac{1}{a^{3}}p\partial_{a}(\Psi_{p}+\Psi_{\pi})
\end{eqnarray}\\
The other two junction conditions are constraints on geometric
potentials and background pentad variables at the position of the
brane and we use them to bring the other four equations into gauge
invariant form. These equation act as boundary conditions for the
bulk perturbation equations.
\section{Conclusion and Discussion}
Gauge properties and symmetries of the teleparallel theory of
gravity is essentially different from those of general relativity.
As a result, one should be careful when defining perturbations in a
manifold with absolute parallelism. The difference in definitions of
perturbations in teleparallel and GR, stems from the fact that when
working with the Stewart lemma, the Lie derivative is different in
two theories. Moreover pentad perturbations in teleparallel gravity
can only contain up to vector modes. These items also implicate the
cosmological perturbations in 4D teleparallel set-up. In 4
dimensions it has been shown that the results arisen by the
perturbed FRW vierbein are different from the ones by perturbed FRW
metric \cite{BA}. This may bring out interests to study some
cosmological issues to find out more about the conceptual role of
torsion in such theories.\\

The five dimensional case is more complicated. Because of the
different junction conditions in this model compared with GR, the
results obtained on the brane cannot be retrieved by general
relativity set-up. In the issue of inflation it has been shown that
in RS model of TEGR, the inflation index grows faster \cite{BAN2}.
Considering these points, studying cosmological perturbations of
this model could be interesting.In this paper, we presented the
fully gauge invariant cosmological perturbation equations for scalar
perturbations in teleparallel gravity by writing down the equations
in terms of the gauge invariant geometric and matter potentials.
These potentials are the teleparallel versions of Bardeen's
potentials in five dimensions. The resulting system of equations can
be solved to uniquely determine the physical scalar perturbation
modes. According to the result obtained in this paper, studying the
cosmological issues in this setup will lead to different conceptual
interpretations and also observational predictions.
\newpage
\appendix{}
\section{}
Non-zero components of the torsion
\begin{eqnarray}
T_{\,\,\,\beta\gamma}^{\alpha}&=&\partial_{\beta}D_{\,\,\,\gamma}^{\alpha}-\partial_{\gamma}D_{\,\,\,\beta}^{\alpha}\nonumber\\
T_{\,\,\,\alpha\beta}^{0}&=&\frac{a}{n}\,\Big(\partial_{\alpha}C_{\beta}-\partial_{\beta}C_{\alpha}\Big)\nonumber\\
T_{\,\,\,50}^{0}&=&\frac{n'}{n}+A'\nonumber\\
T_{\,\,\,\alpha\beta}^{5}&=&a(\partial_{\alpha}F_{\beta}-\partial_{\beta}F_{\alpha})\nonumber\\
T_{\,\,\,5\alpha}^{5}&=&aF'_{\alpha}-\partial_{\alpha}\Phi\nonumber\\
T_{\,\,\,\beta 0}^{\alpha}&=&-\frac{\dot{a}}{a}\delta_{\beta}^{\alpha}+\partial_{\beta}B^{\alpha}-\dot{D}_{\,\,\,\beta}^{\alpha}\nonumber\\
T_{\,\,\,0\alpha}^{0}&=&(\frac{\dot{a}}{n}-\frac{\dot{a}a}{n})C_{\alpha}+\frac{a}{n}\dot{C}_{\alpha}-\partial_{\alpha}A\nonumber\\
T_{\,\,\,5\alpha}^{0}&=&(\frac{a'}{n}-\frac{a'a}{n})C_{\alpha}+\frac{a}{n}C'_{\alpha}\nonumber\\
T_{\,\,\,\alpha 0}^{5}&=&n\partial_{\alpha}\alpha-a\dot{F}_{\alpha}\nonumber\\
T_{\,\,\,05}^{5}&=&\dot{\Phi}-n\alpha'\nonumber\\
T_{\,\,\,50}^{\alpha}&=&(\frac{a'}{a}-\frac{an'}{n})B^{\alpha}+B^{'\alpha}\nonumber\\
T_{\,\,\,\beta 5}^{\alpha}&=&-D_{\,\,\,\,\beta}^{'\alpha}-\frac{a'}{a}\delta_{\beta}^{\alpha}\nonumber\\
\end{eqnarray}

Contorsion
\begin{eqnarray}
K_{\,\,\,\,\,\,\alpha}^{\beta\gamma}&=&-\frac{1}{a^2}\partial^{[\gamma}D_{\,\,\,\alpha]}^{\beta}+\frac{1}{a^2}\partial^{[\beta}D_{~~\alpha]}^{\gamma}
+\frac{1}{a^2}\partial^{[\beta}D_{\alpha}^{\,\,\,\gamma]}-2\frac{\dot{a}}{an^{2}}\delta^{\,\,\,[\beta}_{\alpha}B^{\gamma]}
+2\frac{\dot{a}}{na^{2}}\delta^{\,\,\,[\beta}_{\alpha}C^{\gamma]}\nonumber\\
&~&+2\frac{a'}{a}\delta^{\,\,\,[\beta}_{\alpha}F^{\gamma]}\nonumber\\
K_{\,\,\,\,\alpha}^{0\gamma}&=&-\frac{1}{an}\partial^{[\gamma}C_{\alpha]}+\frac{1}{n^{2}}\Big(\partial_{(\alpha}B^{\gamma)}
+\frac{a'n}{a}\delta^{\gamma}_{\alpha}\alpha-\frac{\dot{a}}{a}\delta^{\gamma}_{\alpha}(1-2A)-\partial_{0}D_{\,\,\,\,\,\alpha)}^{(\gamma}\Big)\nonumber\\
K_{\,\,\,\,\alpha}^{5\gamma}&=&-\frac{1}{a}\partial^{[\gamma}F_{\alpha]}+(\frac{\dot{a}}{an}\alpha+\frac{a'}{a}-2\frac{a'}{a}\Phi)
\delta^{\gamma}_{\alpha}+D_{~~\,\,\,\alpha)}^{'(\gamma}\nonumber\\
K_{\,\,\,\,\alpha}^{5\,0}&=&\frac{1}{2}\Big[-\frac{1}{n}\partial_{\alpha}\alpha-\frac{\dot{a}}{n^{2}}F_{\alpha}+\frac{a}{n^{2}}\dot{F}_{\alpha}
+(\frac{an'}{n^{2}}-\frac{aa'}{n})C_{\alpha}+\frac{a}{n}C'_{\alpha}\nonumber\\
&~&~~~~+(\frac{a^{3}n'}{n^{3}}-\frac{n'a^{2}}{n^{3}})B_{\alpha}
-\frac{a^{2}}{n^{2}}B'_{\alpha}\Big]\nonumber\\
K_{\,\,\,\,0}^{\beta\gamma}&=&-\frac{1}{a^{2}}\partial^{[\gamma}B^{\beta]}+\frac{1}{a^{2}}\partial_{0}D^{[\beta\gamma]}
-\frac{n}{a^{3}}\partial^{[\beta}C^{\gamma]}\nonumber\\
K_{\,\,\,\,0}^{0\gamma}&=&-\frac{1}{a^{2}}\partial^{\gamma}A+(\frac{\dot{a}}{2n^{2}a}-\frac{\dot{a}}{an}+\frac{3\dot{a}}{2a^{2}n})C^{\gamma}
+\frac{n'}{an}F^{\gamma}
-\frac{\dot{a}}{an^{2}}B^{\gamma}+\frac{1}{an}\dot{C}^{\gamma}\nonumber\\
K_{\,\,\,\,0}^{5\gamma}&=&-\frac{1}{2}\Big[(\frac{n'}{n}+\frac{an'}{n}-\frac{2a'}{a})B^{\gamma}-B^{'\gamma}
+(\frac{2a'n}{a^{2}}-\frac{n'}{a}-\frac{a'n}{a})C^{\gamma}+\frac{n}{a}C^{'\gamma}\nonumber\\
&~&+\frac{n}{a^{2}}\partial^{\gamma}\alpha-\frac{\dot{a}}{a^{2}}F^{\gamma}
-\frac{1}{a}\dot{F}^{\gamma}\Big]\nonumber\\
K_{\,\,\,\,0}^{5\,0}&=&\frac{n'}{n}(1-2\Phi)+A'\nonumber\\
K_{\,\,\,\,5}^{\beta\gamma}&=&\frac{1}{a^{2}}D^{'[\beta\gamma]}+\frac{1}{a^{3}}\partial^{[\beta}F^{\gamma]}\nonumber\\
K_{\,\,\,\,5}^{0\gamma}&=&\frac{1}{2}\Big(\frac{1}{an}C'^{\gamma}+(\frac{n'}{n^{3}}-\frac{an'}{n^{3}})B^{\gamma}
+\frac{1}{n^{2}}B^{'\gamma}-(\frac{n'}{an^{2}}-\frac{a'}{na^{2}})C^{\gamma}\Big)\nonumber\\
&~&-\frac{1}{2}\frac{\dot{a}}{a^{2}n^{2}}F^{\gamma}
+\frac{1}{2a^{2}n}\partial^{\gamma}\alpha-\frac{1}{2an^{2}}\dot{F}^{\gamma}\nonumber\\
K_{\,\,\,\,5}^{5\gamma}&=&-\frac{1}{a^{2}}\partial^{\gamma}\Phi+\frac{a'}{a^{2}}F^{\gamma}+\frac{1}{a}F^{'\gamma}
\nonumber\\
K_{\,\,\,\,5}^{5\,0}&=&\frac{\dot{\Phi}}{n^{2}}-\frac{\alpha'}{n}-\frac{n'\alpha}{n^{2}}
\end{eqnarray}
\\
Superpotential
\begin{eqnarray}
S^{~\beta\gamma}_{\alpha}&=&-\frac{1}{a^2}\partial^{[\gamma}D_{\,\,\,\alpha]}^{\beta}+\frac{1}{a^2}\partial^{[\beta}D_{~~\alpha]}^{\gamma}
+\frac{1}{2a^2}\partial^{[\beta}D_{\alpha}^{\,\,\,\gamma]}-(\frac{3\dot{a}}{na^{2}}-\frac{\dot{a}}{an})
\delta^{\,\,\,[\beta}_{\alpha}C^{\gamma]}\nonumber\\
&~&+\frac{1}{a^{2}}\delta^{\,\,\,[\beta}_{\alpha}\partial^{\gamma]}A+\frac{1}{a^{2}}\delta^{\,\,\,[\beta}_{\alpha}\partial^{\gamma]}\Phi
+\frac{1}{a^{2}}\delta^{\beta}_{\alpha}\partial^{[\gamma}D_{~\eta]}^{\eta}
-\frac{1}{a^{2}}\delta^{\gamma}_{\alpha}\partial^{[\beta}D_{~\eta]}^{\eta}\nonumber\\
&~&-\frac{1}{na}\delta^{\,\,\,[\beta}_{\alpha}\dot{C}^{\gamma]}
+\frac{2\dot{a}}{an^{2}}\delta^{\,\,\,[\beta}_{\alpha}B^{\gamma]}
-(\frac{2a'}{a^{2}}+\frac{n'}{an})\delta^{\,\,\,[\beta}_{\alpha}F^{\gamma]}-\frac{1}{a}\delta^{\,\,\,[\beta}_{\alpha}F^{'\gamma]}\nonumber\\
S^{~0\gamma}_{\alpha}&=&\frac{1}{2}\Big[-\frac{1}{an}\partial^{[\gamma}C_{\alpha]}+\frac{1}{n^{2}}(\partial_{(\alpha}B^{\gamma)}
-\delta^{\gamma}_{\alpha}\partial_{\eta}B^{\eta})-\frac{1}{n^{2}}(\partial_{0}D_{~~\alpha)}^{(\gamma}
-\delta^{\gamma}_{\alpha}\partial_{0}D_{~\beta}^{\beta})\nonumber\\
&~&+\Big(\frac{2\dot{a}}{an^{2}}(1-2A)
+\frac{\dot{\Phi}}{n^{2}}-\frac{\alpha}{n}(\frac{2a'}{a}+\frac{n'}{n})-\frac{\alpha'}{n}\Big)\delta^{\gamma}_{\alpha}\Big]\nonumber\\
S^{~5\gamma}_{\alpha}&=&-\frac{1}{2a}\partial^{[\gamma}F_{\alpha]}+\frac{1}{2}(D_{~~~\alpha)}^{'(\gamma}
-\delta^{\gamma}_{\alpha}D_{~~\beta}^{'\beta})-\frac{1}{2}\Big(\frac{2a'}{a}+\frac{n'}{n}+A'+\frac{2\dot{a}}{an}\alpha
-(\frac{4a'}{a}+\frac{2n'}{n})\Phi\Big)\delta^{\gamma}_{\alpha}\nonumber\\
S^{~5\,0}_{5}&=&-\frac{1}{2n^{2}}\Big[-\partial_{\eta}B^{\eta}+\partial_{0}D_{~\eta}^{\eta}+3\frac{\dot{a}}{a}(1-2A)\Big]
+\frac{3a'}{2an}\alpha\nonumber\\
S^{~0\gamma}_{0}&=&\frac{1}{2}\Bigg[(\frac{\dot{a}}{2an^{2}}-\frac{5\dot{a}}{2a^{2}n})C^{\gamma}-\frac{3a'}{a^{2}}F^{\gamma}
+\frac{2\dot{a}}{an^{2}}B^{\gamma}+\frac{2}{a^{2}}\partial^{[\gamma}D_{~~\eta]}^{\eta}\Bigg]\nonumber\\
S^{~5\,0}_{0}&=&-\frac{1}{2}(\frac{3a'}{a}(1-2\Phi)+D_{~~\alpha}^{'\alpha}-\frac{3\dot{a}}{an}\alpha)\nonumber\\
S^{~\gamma5}_{5}&=&\frac{1}{2a^{2}}(\partial^{\gamma}A+2\partial^{[\gamma}D_{~\alpha]}^{\alpha})+
\frac{3\dot{a}}{2an^{2}}B^{\gamma}-(\frac{2\dot{a}}{a^{2}n}-\frac{\dot{a}}{2an})C^{\gamma}-\frac{1}{2na}\dot{C}^{\gamma}\nonumber\\
&~&-(\frac{2a'}{a^{2}}+\frac{n'}{2an})F^{\gamma}+\frac{1}{a^{2}}\partial^{\gamma}\Phi-\frac{1}{a}F^{'\gamma}\nonumber\\
S^{~5\,0}_{\alpha}&=&\frac{1}{2}K_{~~\alpha}^{5\,0}\nonumber\\
S^{~\beta\gamma}_{0}&=&\frac{1}{2}K_{~~0}^{\beta\gamma}\nonumber\\
S^{~5\gamma}_{0}&=&\frac{1}{2}K_{~~0}^{5\gamma}\nonumber\\
S^{~\beta\gamma}_{5}&=&\frac{1}{2}K_{~~5}^{\beta\gamma}\nonumber\\
S^{~0\gamma}_{5}&=&\frac{1}{2}K_{~~5}^{0\gamma}\nonumber\\
\end{eqnarray}
\\
Torsion scalar
\begin{eqnarray}
T&=&\frac{6\dot{a}^{2}}{a^{2}n^{2}}-\frac{6a'}{a}(\frac{a'}{a}+\frac{n'}{n})-\frac{4\dot{a}}{an^{2}}(\partial_{\alpha}
B^{\alpha}-\partial_{0}D_{~~\alpha}^{\alpha})-\frac{12\dot{a}^{2}}{a^{2}n^{2}}A-\frac{6\dot{a}}{a}(\frac{2a'}{an}-\frac{n'}{n^{2}})\alpha\nonumber\\
&~&-\frac{6\dot{a}}{an}\alpha'+\frac{6\dot{a}}{an^{2}}\dot{\Phi}-(\frac{4a'}{a}+\frac{2n'}{n})D_{~~\alpha}^{'\alpha}
+12\frac{a'}{a}(\frac{a'}{a}+\frac{n'}{n})\Phi-6\frac{a'}{a}A'\nonumber\\
\end{eqnarray}
\section{}
In this appendix we present the left hand side of the perturbed
teleparallel field equations in full.
\begin{eqnarray}
F^{\alpha}_{a}&=&\frac{1}{a}\Bigg[\partial_{\beta}\Bigg(\frac{1}{a^{2}}\partial^{[\beta}D_{~~a]}^{\alpha}
-\frac{1}{a^{2}}\partial^{[\alpha}D_{~~a]}^{\beta}
+\frac{1}{2a^{2}}\partial^{[\beta}D_{a}^{~~\alpha]}-(\frac{3\dot{a}}{a^{2}n}-\frac{\dot{a}}{an})\delta^{[\beta}_{a}C^{\alpha]}\nonumber\\
&~&~~~~-(\frac{2a'}{a^{2}}+\frac{n'}{na})\delta^{~[\beta}_{a}F^{\alpha]}
-\frac{1}{a}\delta^{~[\beta}_{a}F^{'\alpha]}+\frac{1}{a^{2}}\delta^{~[\beta}_{a}\partial^{\alpha]}A
+\frac{1}{a^{2}}\delta^{~[\beta}_{a}\partial^{\alpha]}\Phi\nonumber\\
&~&~~~~+\frac{1}{a^{2}}\delta^{~\beta}_{a}\partial^{[\alpha}D_{~~\gamma]}^{\gamma}
-\frac{1}{a^{2}}\delta^{~\alpha}_{a}\partial^{[\beta}D_{~~\gamma]}^{\gamma}-\frac{1}{a}\delta^{~[\beta}_{a}\dot{C}^{\alpha]}
+\frac{2\dot{a}}{an^{2}}\delta^{~[\beta}_{a}B^{\alpha]}\Bigg)\Bigg]\nonumber\\
&~&~~~~+\frac{1}{a^{3}n^{2}}\partial_{0}\Bigg[-\frac{a}{2}\partial^{[\alpha}C_{a]}
+\frac{a^{2}}{2n}(\partial_{(a}B^{\alpha)}-\delta^{\alpha}_{a}\partial_{\beta}B^{\beta})+\frac{\dot{a}a}{n}D_{~~a}^{\alpha}\nonumber\\
&~&~~~~+\frac{\dot{a}a}{n}(1-2A)\delta^{\alpha}_{a}-\frac{a^{2}}{2}(\frac{2a'}{a}+\frac{n'}{n})\alpha\delta^{\alpha}_{a}
-\frac{a^{2}}{2n}(\partial_{0}D_{~~a]}^{[\alpha}-\delta^{\alpha}_{a}
\partial_{0}D_{~~\beta}^{\beta})\nonumber\\
&~&~~~~+(\frac{a^{2}}{2n}\dot{\Phi}-\frac{a^{2}}{2}\alpha')\delta^{\alpha}_{a}\Bigg]
+\frac{\dot{a}}{a^{2}n^{2}}\delta^{\alpha}_{a}\partial_{0}(A+\Phi+|D|)\nonumber\\
&~&~~~~+\frac{1}{a^{3}n}\partial_{5}\Bigg[\frac{-an}{2}\partial^{[\alpha}F_{a]}+\frac{a^{2}n}{2}(D_{~~~a)}^{'(\alpha}
-\delta^{\alpha}_{a}D_{~~\beta}^{'\beta})
+a^{2}n(\frac{a'}{a}+\frac{n'}{2n})D^{\alpha}_{~~a}\nonumber\\
&~&~~~~-(a\dot{a}\alpha+a'n\alpha+\frac{a^{2}n'}{2}-2aa'n\Phi-a^{2}n'\Phi+\frac{a^{2}n}{2}A')\delta^{\alpha}_{a}\Bigg]\nonumber\\
&~&~~~~-(\frac{a'}{a^{2}}+\frac{n'}{2an})\delta^{\alpha}_{a}\partial_{5}(A+\Phi+|D|)+\frac{5}{2}(-\frac{\dot{a}^{2}}{a^{3}n^{2}}+\frac{a^{'2}}{a^{3}}
+\frac{a'n'}{a^{2}n})D^{\alpha}_{~~a}\nonumber\\
&~&~~~~-\frac{\dot{a}}{2a^{2}n^{2}}\Bigg(2\partial_{a}B^{\alpha}-\partial_{(a}B^{\alpha)}+3\delta^{\alpha}_{a}\partial_{\beta}B^{\beta}
-2\partial_{0}D^{\alpha}_{~~a}+\partial_{0}D_{~~a)}^{(\alpha}-3\delta^{\alpha}_{a}\partial_{0}D_{~~\beta}^{\beta})\Bigg)\nonumber\\
&~&~~~~+\Bigg(\frac{5}{2}\frac{\dot{a}^{2}}{a^{3}n^{2}}-\frac{5\dot{a}^{2}}{a^{3}n^{2}}A-\frac{5a'\dot{a}}{a^{3}n}\alpha-\frac{5a^{'2}}{2a^{3}}
-\frac{2a'n'}{a^{2}n}+\frac{5a^{'2}}{a^{3}}\Phi+\frac{4a'n'}{a^{2}n}\Phi-\frac{2a'}{a^{2}}A'\nonumber\\
&~&~~~~-\frac{3a'}{2a^{2}}D_{~\beta}^{'\beta}+\frac{\dot{a}n'}{a^{2}n^{2}}\alpha-\frac{2\dot{a}}{a^{2}n}\alpha'
+\frac{\dot{a}}{a^{2}n^{2}}\dot{\Phi}\Bigg)\delta^{\alpha}_{a}-\frac{\dot{a}}{2a^{3}n}\partial^{[\alpha}C_{a]}
-\frac{a'}{2a^{3}}\partial^{[\alpha}F_{a]}\nonumber\\
&~&~~~~-\frac{n'}{2na}D_{~~\beta}^{'\beta}\delta^{\alpha}_{a}
\nonumber\\
\end{eqnarray}
\\
\begin{eqnarray}
F^{0}_{\bar{0}}&=&\partial_{\beta}\Bigg[-\frac{1}{2n}(\frac{\dot{a}}{2an^{2}}-\frac{5\dot{a}}{2a^{2}n})C^{\beta}
-\frac{1}{2a^{2}n}\partial^{[\beta}D_{~~\gamma]}^{\gamma}
+\frac{3a'}{2a^{2}n}F^{\beta}\Bigg]-\frac{3a'}{2an}\partial_{5}(A+\Phi+|D|)\nonumber\\
&~&+\frac{1}{a^{3}n}\partial_{5}\Bigg(\frac{a^{3}}{2}D_{~~\beta}^{'\beta}-\frac{3a'a^{2}}{2}(1-2\Phi)+\frac{3a'a^{2}}{2}A\Bigg)
+\frac{5}{2}\frac{\dot{a}^{2}}{a^{2}n^{3}}-3\frac{a'n^{'2}}{an^{2}}\nonumber\\
&~&-\frac{3a'}{2na}A'-(\frac{n'}{n}+\frac{a'}{an})D_{~~\beta}^{'\beta}-(\frac{9\dot{a}^{2}}{2a^{2}n^{3}}
-\frac{3a^{'2}}{2a^{2}n})A+
3(\frac{a^{'2}}{a^{2}n}+\frac{2a'n'}{an^{2}})\Phi\nonumber\\
&~&-\frac{2\dot{a}}{an^{3}}(\partial_{\beta}B^{\beta}-
\partial_{0}D_{~~\beta}^{\beta})+(\frac{\dot{a}n'}{an^{3}}-\frac{3a'\dot{a}}{a^{2}n^{2}}-\frac{3n'\dot{a}}{2an^{2}})\alpha
-\frac{\dot{a}}{2an^{3}}(7n\alpha'+\dot{\Phi})
\end{eqnarray}
\\
\begin{eqnarray}
F^{0}_{\bar{5}}&=&\partial_{\beta}\Bigg[-\frac{1}{4an}C^{'\beta}+\frac{1}{4}(\frac{an'}{n^{3}}-\frac{n'}{n^{3}})B^{\beta}-\frac{1}{4n^{2}}B^{'\beta}
+\frac{1}{4}(\frac{n'}{an^{2}}-\frac{a'}{na^{2}})C^{\beta}+\frac{\dot{a}}{4a^{2}n^{2}}F^{\beta}
\nonumber\\
&~&-\frac{1}{4a^{2}n}\alpha+\frac{1}{4an^{2}}\dot{F}^{\beta}\Bigg]-\frac{3\dot{a}}{2an^{2}}\partial_{5}(A+\Phi+|D|)+\frac{\dot{a}a'}{a^{2}n^{2}}
(1-2A)\nonumber\\
&~&+\frac{1}{a^{3}n}\partial_{5}\Bigg[\frac{a^{3}}{2n}(\partial_{\beta}B^{\beta}-\partial_{0}D_{~~\beta}^{\beta})
-\frac{3\dot{a}a^{2}}{2n}(1-2A)+\frac{3a'a^{2}}{2}\alpha+\frac{3\dot{a}a^{2}}{2n}\Phi\Bigg]\nonumber\\
&~&-\frac{a'}{2an}(\frac{2a'}{a}+\frac{n'}{n})\alpha+\frac{a'}{2an^{2}}\dot{\Phi}-\frac{a'}{2an}\alpha'
-\frac{\dot{a}}{an^{2}}(D_{~~\alpha}^{'\alpha}+\Phi)
\end{eqnarray}
\\
\begin{eqnarray}
F^{0}_{a}&=&\partial_{\beta}\Bigg[\frac{a}{2}\partial^{[\beta}C_{a]}+\frac{a^{2}}{2n}(\partial_{(a}B^{\beta)}
-\delta^{\beta}_{a}\partial_{\gamma}B^{\gamma})-\frac{a^{2}}{2n}\dot{\Phi}\delta^{\beta}_{a}+\frac{a^{2}}{2n}(\partial_{0}D_{~~a)}^{(\beta}-
\delta^{\beta}_{a}\partial_{0}D_{~~\gamma}^{\gamma})\nonumber\\
&~&-\frac{\dot{a}a}{n}(1-2A)\delta^{\beta}_{a}+(\frac{a'a}{2}+\frac{n'a^{2}}{2n})\alpha\delta^{\beta}_{a}+\frac{a^{2}}{2n}\alpha'\delta^{\beta}_{a}\Bigg]
-\frac{\dot{a}a}{n}\partial_{a}(\Phi+A+|D|)\nonumber\\
&~&+\frac{1}{a^{3}n}\partial_{5}\Bigg[-\frac{a^{2}}{4}\partial_{a}\alpha+\frac{a^{3}}{4n}\dot{F}_{a}
+(\frac{a^{3}n'}{4n}+\frac{5a'a^{3}}{4})C_{a}+\frac{a^{3}}{4}C'_{a}+\frac{5\dot{a}a^{2}}{4n}F_{a}-\frac{a^{4}}{4n}B'_{a}\nonumber\\
&~&+\frac{a^{4}n'}{4n^{2}}(a-1)B_{a}\Bigg]-\frac{2\dot{a}}{a^{2}n^{2}}\partial_{\,[\gamma}D_{~~a]}^{\gamma}
+(-\frac{9\dot{a}^{2}}{2an^{3}}-\frac{3a^{'2}}{a^{2}n}+\frac{11a^{'2}}{4an}+\frac{a'n'}{4an^{2}}+\frac{3a'n'}{2n^{2}})C_{a}\nonumber\\
&~&+\frac{3\dot{a}}{2a^{2}n^{2}}\partial_{a}\Phi-\frac{13\dot{a}a'}{4a^{2}n^{2}}F_{a}-\frac{3\dot{a}}{2an^{2}}F'_{a}
+\frac{a'}{4an^{2}}\dot{F}_{a}-\frac{5}{4}\frac{a'}{an}C'_{a}+\frac{a'n'}{4n^{3}}(a-1)B_{a}\nonumber\\
&~&-\frac{a'}{4a^{2}n}\partial_{i}\alpha-\frac{a'}{4n^{2}}B'_{a}\nonumber\\
\end{eqnarray}
\\
\begin{eqnarray}
F^{5}_{\bar{5}}&=&\partial_{\beta}\Bigg[-(\frac{2a'}{a^{2}}+\frac{n'}{2an})F^{\beta}+\frac{1}{a^{2}}\partial^{\beta}\Phi-\frac{1}{a}F^{'\beta}
+\frac{1}{a^{2}}\partial^{[\beta}D_{~~\gamma]}^{\gamma}+\frac{1}{2a^{2}}\partial^{\beta}A\nonumber\\
&~&+\frac{3\dot{a}}{2an^{2}}B^{\beta}-(\frac{2\dot{a}}{a^{2}n}-\frac{\dot{a}}{2an})C^{\beta}-\frac{1}{2an}\dot{C}^{\beta}\Bigg]
+\frac{3\dot{a}}{2an^{2}}\partial_{0}(A+\Phi+|D|)\nonumber\\
&~&+\frac{1}{a^{3}n}\partial_{0}\Bigg[-\frac{a^{3}}{2n}(\partial_{\beta}B^{\beta}
-\partial_{0}D_{~~\gamma}^{\gamma})+\frac{3\dot{a}a^{2}}{2n}(1-A-\Phi)-\frac{3a'a^{2}}{2}\alpha\Bigg]\nonumber\\
&~&-\frac{5a^{'2}}{2a^{2}}-\frac{7a'n'}{2an}-\frac{7a'}{2a}A'-(\frac{3a'}{2a}+\frac{n'}{n})D_{~~\gamma}^{'\gamma}
+\frac{3\dot{a}}{an^{2}}\dot{\Phi}+\frac{3\dot{a}^{2}}{2a^{2}n^{2}}-\frac{3\dot{a}}{an}\alpha'\nonumber\\
&~&-\frac{\dot{a}}{an^{2}}(\partial_{\beta}B^{\beta}-\partial_{0}D_{~~\gamma}^{\gamma})
+(\frac{5a'^{2}}{2a^{2}}
+\frac{a'n'}{2an}+\frac{3\dot{a}^{2}}{2a^{2}n^{2}})\Phi\nonumber\\
&~&+(\frac{3\dot{a}a'}{2an^{2}}-\frac{4\dot{a}a'}{a^{2}n})\alpha
\end{eqnarray}
\\
\begin{eqnarray}
F^{5}_{a}&=&(\frac{a'}{a^{2}}+\frac{n'}{2na})\partial_{a}(\Phi+A+|D|)+\partial_{\beta}\Bigg[\frac{1}{2a^{2}}\partial^{[\beta}F_{a]}
-\frac{1}{2a}(D_{~~~a)}^{'(\beta}-\delta_{a}^{\beta}D_{~~\gamma}^{'\gamma})\nonumber\\
&~&+\Big(\frac{a'}{a^{2}}+\frac{n'}{2an}+\frac{1}{2a}A'-(\frac{2a'}{a^{2}}+\frac{n'}{an})\Phi+\frac{\dot{a}}{a^{2}n^{2}}\alpha\Big)\delta_{a}^{\beta}
-\frac{1}{2a}(\frac{2a'}{a}+\frac{n'}{n})D_{~~a}^{\beta}\Bigg]\nonumber\\
&~&+\frac{1}{a^{3}n}\partial_{0}\Bigg[-\frac{5a^{2}\dot{a}}{4n}F_{a}-\frac{a^{3}}{4n}\dot{F}_{a}-(\frac{a^{3}n'}{4n}-\frac{5a'a^{3}}{4})C_{a}
-\frac{a^{3}}{4}C'_{a}-\frac{a^{4}n'}{4n^{2}}(a-1)B_{a}\nonumber\\
&~&-\frac{a^{4}}{4n}B'_{a}+\frac{a^{2}}{4}\partial_{a}\alpha\Bigg]+(\frac{2a'}{a^{2}}
+\frac{n'}{an})\partial_{\,[\beta}D_{~~a]}^{\beta}+\frac{3a'}{2an}\dot{C}_{a}-\frac{3a'}{2a^{2}}\partial_{a}A-\frac{\dot{a}}{4an}C'_{a}\nonumber\\
&~&+\frac{\dot{a}n'}{4n^{3}}(a-1)B_{a}+\frac{\dot{a}}{4n^{2}}B'_{a}+(\frac{5a^{'2}}{2a^{2}}+\frac{7a'n'}{2an}-\frac{5\dot{a}^{2}}{4a^{2}n^{2}})F_{a}
-\frac{5\dot{a}}{4a^{2}n}\partial_{a}\alpha+\frac{5\dot{a}}{4an^{2}}\dot{F}_{a}\nonumber\\
&~&+\Bigg(-\frac{3\dot{a}a'}{4an}+\frac{\dot{a}n'}{2n^{2}}+\frac{3a'\dot{a}}{a^{2}n}-\frac{\dot{a}n'}{4an^{2}}\Bigg)C_{a}\nonumber\\
\end{eqnarray}
\\
\begin{eqnarray}
F^{\alpha}_{\bar{5}}&=&\partial_{\beta}\Bigg[\frac{1}{2a^{3}}\partial^{\,[\beta}F^{\alpha]}+\frac{1}{2a^{2}}D^{'[\beta\alpha]}
\Bigg]+\frac{1}{a^{3}n}\partial_{5}\Bigg[\frac{an}{2}\partial^{\alpha}A-an\partial^{[\alpha}D_{~~\beta]}^{\beta}
-\frac{3\dot{a}a^{2}}{2n}B^{\alpha}-an\partial^{\alpha}\Phi\nonumber\\
&~&+(2\dot{a}a-\frac{\dot{a}a^{2}}{2})C^{\alpha}+\frac{a^{2}}{2}\dot{C}^{\alpha}+(2a'an+\frac{a^{2}n'}{2})F^{\alpha}
+a^{2}nF^{'\alpha}\Bigg]+\frac{1}{a^{3}n}\partial_{0}\Bigg[\frac{a^{2}}{4}C^{'\alpha}\nonumber\\
&~&+(\frac{a^{3}n'}{4n^{2}}-\frac{a^{4}n'}{4n^{2}})B^{\alpha}+\frac{a^{3}}{4n}B^{'\alpha}
-(\frac{a^{2}n'}{4n}-\frac{a'a}{4})C^{\alpha}-\frac{\dot{a}a}{4n}F^{\alpha}-\frac{a^{2}}{4n}\dot{F}^{\alpha}
+\frac{a}{4}\partial^{\alpha}\alpha\Bigg]\nonumber\\
&~&-\frac{\dot{a}a'}{a^{2}n^{2}}B^{\alpha}+(\frac{a'}{a^{3}}-\frac{n'}{a^{2}n})\partial^{[\alpha}D_{~~\beta]}^{\beta}
-\frac{\dot{a}}{an^{2}}B^{'\alpha}-(\frac{\dot{a}n'}{4an^{3}}-\frac{5\dot{a}n'}{4a^{2}n^{2}})C^{\alpha}+\frac{3a'n'}{2na^{2}}F^{\alpha}\nonumber\\
\end{eqnarray}
\\
\begin{eqnarray}
F^{\alpha}_{\bar{0}}&=&\partial_{\beta}\Bigg[-\frac{1}{2na^{2}}(\partial^{[\alpha}B^{\beta]}-\partial_{0}D^{[\beta\alpha]})
-\frac{1}{2a^{3}}\partial^{[\beta}C^{\alpha]}\Bigg]\nonumber\\
&~&+\frac{1}{a^{3}n}\partial_{0}\Bigg[\frac{\dot{a}a^{2}}{n^{2}}(1-a)B^{\alpha}+(\frac{\dot{a}a^{2}}{4n^{2}}-\frac{5\dot{a}a}{4n})C^{\alpha}
-\frac{3aa'}{2}F^{\alpha}+a\partial^{[\alpha}D_{~~\beta]}^{\beta}\Bigg]\nonumber\\
&~&+\frac{1}{a^{3}n}\partial_{5}\Bigg[(a'a^{3}+\frac{n'a^{4}}{2n}-\frac{3\dot{a}a}{n})B^{\alpha}+(2\dot{a}a-\frac{\dot{a}a^{2}}{2})C^{\alpha}
+\frac{a^{2}}{2}\dot{C}^{\alpha}-an\partial^{\alpha}\Phi-\frac{an}{2}\partial^{\alpha}A\nonumber\\
&~&+(2a'an+\frac{a^{2}n'}{2})F^{\alpha}+a^{2}n
F^{'\alpha}-an\partial^{\,[\alpha}D_{~~\beta]}^{\beta}\Bigg)\Bigg]
+\frac{3\dot{a}}{2na^{3}}\partial^{[\alpha}D_{~~\beta]}^{\beta}-\frac{a'}{an}B^{'\alpha}\nonumber\\
&~&-\frac{n'}{2a^{2}n}\partial^{\alpha}\alpha+\frac{n'}{2an^{2}}\dot{F}^{\alpha}+\frac{n'\dot{a}}{2n^{2}a^{3}}F^{\alpha}-\frac{n'}{2an}C^{'\alpha}
\nonumber\\
&~&+\Big(-\frac{n^{'2}}{2n^{3}}-\frac{5\dot{a}}{2an^{3}}+\frac{5a^{'2}}{2an}+\frac{2a'n'}{n^{2}}-\frac{a^{'2}}{a^{2}n}
+\frac{n'a'}{2an^{2}}+\frac{an^{'2}}{2n^{3}}\Big)B^{\alpha}\nonumber\\
\end{eqnarray}
Where $|D|$ is the determinant of $D^{i}_{~\mu}$.

\end{document}